\documentclass[fleqn,twoside]{article}
\usepackage{espcrc2}
\usepackage{graphicx}

\newcommand{\ket}[1]{\vert {#1} \rangle }

\newcommand{\ie}{{\it i.e.}}
\newcommand{\eg}{{\it e.g.}}
\renewcommand{\bar}[1]{\overline{#1}}

\newcommand{\qu}{{\rm q}}

\newcommand{\ieps}{i\varepsilon}

\newcommand{\order}[1]{${ O}\left(#1 \right)$}

\newcommand{\beq}{\begin{equation}}
\newcommand{\eeq}{\end{equation}}
\newcommand{\beqa}{\begin{eqnarray}}
\newcommand{\eeqa}{\end{eqnarray}}
\newcommand{\VEV}[1]{\left\langle{#1}\right\rangle}
\newcommand{\etal}{{\em et al.}}

\title{ Physics at the Light-Front }

\author{S. J. Brodsky\thanks{Work supported by the Department of Energy
under contract number DE--AC03--76SF00515.} \address{Stanford Linear
Accelerator Center,
 2575 Sand Hill Road, Menlo Park, CA 94025}}

\begin{document}

\begin{abstract}
The light-front representation of quantum chromodynamics provides a
frame-independent, quantum-mechanical representation of hadrons at the
amplitude level, capable of encoding their multi-quark, hidden-color and gluon
momentum, helicity, and flavor correlations in the form of universal
process-independent hadron wavefunctions.
The universality and frame-independence of the LCWF's thus allow a
profound connection between diffractive dissociation,  hard
scattering exclusive processes such as elastic form factors,
two-photon reactions, and heavy hadron decays.  In this concluding talk of
the ECT*
International Conference On Light-Cone Physics: Particles And Strings (Trento
2001), I review recent calculations and new applications of light-front
wavefunctions in QCD and other theories.  I also review the
distinction between the structure functions measured in deep inelastic lepton
scattering and the quark distributions determined from light-front
wavefunctions.
 \vspace{1pc}
\end{abstract}

\maketitle

\section{INTRODUCTION}

Light-front quantization has become a standard tool for the
analysis of nonperturbative problems in quantum field theory as
well as in string~\cite{Metsaev:2000yu} and
$M$-theory~\cite{Susskind:1997cw}.  The light-front method is
especially useful for quantum chromodynamics, since it provides a
physically appealing yet rigorous extension of many-body quantum
mechanics to relativistic bound states: the multi-quark,
hidden-color and gluon momentum, helicity, and flavor correlations
of a hadron are encoded in the form of universal
process-independent Lorentz-invariant wavefunctions.  In higher
dimension theories, the discretization of light-front momenta
provides a way to describe the evolution of the world-sheet of
strings~\cite{Bardakci:2001cn}.

The ECT* International Conference On Light-Cone
Physics: Particles And Strings (Trento 2001) provided an
outstanding forum to assess the continuing progress in this field.  This
report can only
cover a portion of the many interesting developments in light-cone
physics made this past year.  I will also review a number of new
directions and applications.

Quantization at fixed light-cone time $\tau=t+z/c$~\cite{Dirac:cp}
allows one to apply Heisenberg matrix mechanics to relativistic quantum
field theory.  The Heisenberg equation on the light-front,
\begin{equation}
H_{LC} \ket{\Psi} = M^2 \ket{\Psi}\ ,
\end{equation}
can, at least in principle, be solved by diagonalizing the matrix
$\VEV{n|H_{LC}|m}$ on the free Fock basis:~\cite{Brodsky:1997de}
The operator $H_{LC}$ can be derived from the Lagrangian of the theory
using canonical quantization.  The eigenvalues $\{M^2\}$ of
$H_{LC}=H^{0}_{LC} + V_{LC}$ then give the squared invariant masses of the
bound and continuum spectrum of the theory.
The projections $\{\VEV{n|\Psi}\}$ of the eigensolution
on the $n$-particle Fock states provide the light-front wavefunctions.
Thus solving a quantum field theory is equivalent to solving
a coupled many-body quantum mechanics problem:
\begin{equation}
\left[M^2 - \sum_{i=1}^n{m_{\perp i}^2\over x_i}\right] \psi_n =
\sum_{n'}\int \VEV{n|V_{LC}|n'} \psi_{n'}
\label{eq:feqm}
\end{equation}
where the convolution and sum is understood over the Fock number,
transverse momenta, light-cone momentum fractions $x_i =
k^+_i/P^+_\pi = (k^0 + k^z_i)/(P^0+P^z)$ and spin projections of
the intermediate states.  Here $m^2_\perp = m^2 + k^2_\perp.$
Remarkably, the formalism is frame-independent; \ie, $H_{LC}$ and
the light-cone variables are independent of the total $P^+ = P^0 +
P^z$ and $P_\perp$ of the system.  A general review of light-front
quantization methods is given in Ref.~\cite{Brodsky:1997de}.

The light-front wavefunctions $\psi_{n/p}(x_i,\vec k_{\perp
i},\lambda_i)$ of a hadron in QCD are the projections of the
hadronic eigenstate on the free color-singlet Fock state $|n>$ at
a given light-cone time $\tau = t+ z/c.$ They thus represent the
ensemble of quark and gluon states possible when a hadron is
intercepted at the light-front.  The wavefunctions are
Lorentz-invariant functions of the light-front momentum fractions
$x_i$ with $\sum^n_{i=1} x_i = 1$ and ${\vec k_{\perp i}}$ with
$\sum^n_{i=1} {\vec k_{\perp i}} = {\vec 0_\perp}$ and are
independent of the bound state's physical momentum $P^+$, and
$P_\perp$~\cite{Lepage:1980fj}.  The actual physical transverse
momenta are ${\vec p_{\perp i}} = x_i {\vec P_\perp} + {\vec
k_{\perp i}}.$ The $\lambda_i$ label the light-front spin $S^z$
projections of the quarks and gluons along the quantization $z$
direction.  The physical gluon polarization vectors
$\epsilon^\mu(k,\ \lambda = \pm 1)$ are specified in light-cone
gauge by the conditions $k \cdot \epsilon = 0,\ \eta \cdot
\epsilon = \epsilon^+ = 0.$ The parton degrees of freedom are thus
all physical; there are no Faddeev-Popov ghost terms or negative
metric states. The spinors of the light-front formalism
automatically incorporate the Melosh-Wigner rotation.

The wavefunctions derived from light-front quantization play a
central role in QCD phenomenology, providing a frame-independent
description of hadrons in terms of their quark and gluon degrees
of freedom.  A recent summary of their phenomenological
applications can be found in Ref.~\cite{Brodsky:2001dx}.  For
example, any spacelike and timelike form factors and matrix
elements of local currents $\VEV{B|J^\mu(0)|A}$, such as the
semileptonic decay amplitudes of heavy hadrons have an exact
representation as overlap momentum-space overlap integrals of
light-front wavefunctions~\cite{Brodsky:1980zm}.  The light-front
Fock representation is particularly important for semi-leptonic
exclusive matrix elements such as $B \to D \ell \bar \nu$.  The
Lorentz-invariant description requires both the overlap of $n' =
n$ parton-number conserving wavefunctions as well as the overlap
of wavefunctions with parton numbers $n' = n-2$ which arises from
the annihilation of a quark-antiquark pair in the initial
wavefunction~\cite{Brodsky:1999hn}.  The skewed parton
distributions which control deeply virtual Compton scattering can
also be defined from diagonal and off-diagonal overlaps of
light-front wavefunctions~\cite{Brodsky:2000xy}.

The key non-perturbative input for exclusive
processes involving large momentum transfer is the gauge- and
frame-independent hadron distribution amplitude
\cite{Lepage:1979zb,Lepage:1980fj} defined as the integral of the valence
(lowest particle number) Fock wavefunction;
\eg\ for the pion
$
\phi_\pi (x_i,\Lambda) \equiv \int d^2k_\perp\, \psi^{(\Lambda)}_{q\bar
q/\pi} (x_i, \vec k_{\perp i},\lambda)
$
where the global cutoff $\Lambda$ is identified with the
resolution $Q$.  The distribution amplitude controls leading-twist
exclusive amplitudes at high momentum transfer, and it can be
related to the gauge-invariant Bethe-Salpeter wavefunction at
equal light-cone time.  The logarithmic evolution of hadron
distribution amplitudes $\phi_H (x_i,Q)$ can be derived from the
perturbatively-computable tail of the valence light-front
wavefunction in the high transverse momentum regime
\cite{Lepage:1979zb,Lepage:1980fj}.  The conformal basis for the
evolution of the three-quark distribution amplitudes for the
baryons~\cite{Lepage:1979za} has been obtained by Braun \etal \
\cite{Braun:1999te}.
It is particularly important to understand the shape of the gauge-
and process-independent meson and baryon valence-quark
distribution amplitudes~\cite{Lepage:1980fj} $\phi_M(x,Q)$, and
$\phi_B(x_i,Q)$.  These quantities specify how a hadron shares its
longitudinal momentum among its valence quarks; they control
virtually all exclusive processes involving a hard scale $Q$,
including form factors, Compton scattering, and photoproduction at
large momentum transfer.  New applications of the perturbative QCD
factorization formalism were presented at this meeting in
Ref.~\cite{Choi:2001ew}.

Progress in measuring the basic parameters of electroweak
interactions and $CP$ violation will require a quantitative
understanding of the dynamics and phase structure of $B$ decays at
the amplitude level.  Factorization theorems have now been proven
which allow one to rigorously compute certain types of exclusive
$B$ decays in terms of the light-front wavefunctions and
distribution amplitudes of B meson and the final state
hadrons~\cite{Beneke:1999br}.  Recent progress in this field was
reviewed at this meeting by Li~\cite{Li:2001vm}.

\section{STRUCTURE FUNCTIONS VERSUS LIGHT-FRONT WAVEFUNCTIONS}

The quark and gluon distributions of hadrons, including the spin
and transversity distributions which enter hard inclusive
reactions in QCD, can be defined as probability measures and density
matrices of the light-front wavefunctions.  For example, the
quark distribution in a hadron $H$ is
\begin{eqnarray}
&& \!\!\!\!\!\!\!\!{P}_{\qu/H}(x_{Bj},Q^2) = \sum_n
\int^{k_{i\perp}^2<Q^2}
\left[
\prod_i\, dx_i\, d^2k_{\perp i}\right]\nonumber \\
&&\times |\psi_{n/H}(x_i,k_{\perp i})|^2 \sum_{j=q}
\delta(x_{Bj}-x_j).
\end{eqnarray}
It has been conventional to identify the leading-twist structure
functions $F_i(x,Q^2)$ measured in deep inelastic lepton
scattering with the light-front probability distributions.  For
example, in the parton model,
$F_2(x,Q^2) = \sum_q e^2_q x {P}_{\qu/H}(x,Q^2).$ However, Hoyer,
Marchal, Peigne, Sannino, and I~\cite{Brodsky:2001ue} have
recently shown that the leading-twist contribution to deep
inelastic scattering is affected by diffractive rescattering of a
quark in the target, a coherent effect which is not included in
the light-front wavefunctions, even in light-cone gauge.  The effective gluon
propagator in light-cone gauge $A^+=0$ is singular:
\begin{equation}
d_{LC}^{\mu\nu}(k) =
\frac{i}{k^2+\ieps}\left[-g^{\mu\nu}+\frac{n^\mu k^\nu+ k^\mu
n^\nu}{n\cdot k}\right] .\label{lcprop}
\end{equation}
It has a pole at $k^+ \equiv n\cdot k = 0,$ which can be defined
by an analytic prescription such as the Mandelstam-Liebbrandt
prescription~\cite{Leibbrandt:1987qv}.  In final-state scattering
involving on-shell intermediate states, the exchanged momentum
$k^+$ is of \order{1/\nu} in the target rest frame, which enhances
the second term of the light-cone gauge propagator.  This
enhancement allows rescattering to contribute at leading twist
even in LC gauge.  We have verified in Feynman and light-cone
gauge that diffractive contributions to the deep inelastic
scattering $\gamma^* p \to X p^\prime$ cross sections, which leave
the target intact, contribute at leading twist to deep inelastic
scattering~\cite{Brodsky:2001ue}.

Diffractive events resolve the quark-gluon structure of the
virtual photon, not the quark-gluon structure of the target, and
thus they give contributions to structure functions which are not
target parton probabilities.  Our analysis of deep inelastic
scattering $\gamma^*(q) p \to X$, when interpreted in frames with
$q^+ > 0,$ also supports the color dipole description of deep
inelastic lepton scattering at small $x_{bj}$.  For example, in
the case of the aligned-jet configurations, one can understand
$\sigma_T(\gamma^* p)$ at high energies as due to the coherent
color gauge interactions of the incoming quark-pair state of the
photon interacting, first coherently and finally incoherently, in
the target. The distinction between structure functions and target
parton probabilities is also implied by the Glauber-Gribov picture
of nuclear shadowing~\cite{Gribov:1969jf}. In this framework,
shadowing arises from interference of rescattering amplitudes
involving diffraction channels and on-shell intermediate states.
In contrast, the wave function of a stable target is strictly real
since it does not have on energy-shell configurations.  Thus
nuclear shadowing is not a property of the light-front
wavefunctions of a nuclear target; rather, it involves the total
dynamics of the $\gamma^*$-nucleus collision.  A strictly
probabilistic interpretation of the deep inelastic cross section
cross section is thus precluded.

\section{COLOR TRANSPARENCY, DIF\-FRACTIVE PROCESSES,
AND LIGHT-FRONT WAVEFUNCTIONS}

One of the features of QCD which distinguishes it from
traditional hadron physics is the fact that hadrons have interaction cross
sections
which fluctuate according to the size of its color dipole moment.
Thus valence Fock states with small impact separation
between the constituents will interact weakly and thus can
transverse a nucleus with minimal interactions.
At high
energies the Fock states of a hadron with small transverse size
interact weakly even in a nuclear target because of their small
dipole moment.  This is the basis of ``color transparency'' in
perturbative QCD~\cite{Brodsky:1988xz,Bertsch:1981py}.
The
amplitude for the diffractive dissociation of a hadron into jets at high
energies is given by a transverse momentum derivative of its
light-front wavefunction, so it is possible to measure the
wavefunctions of a relativistic hadron by diffractively dissociating it
into jets whose momentum distribution is correlated with the valence
quarks' momenta
\cite{Ashery:1999nq,Bertsch:1981py,Frankfurt:1993it,Frankfurt:2000tq}.
Photon
exchange measures a weighted sum of transverse derivatives
$\partial_{k_\perp} \psi_n(x_i, k_{\perp_i},\lambda_i),$ and two-gluon
exchange measures the second transverse partial derivative~\cite{BHDP}.
The diffractive dijet dissociation experiment E791 at
Fermilab using 500 GeV incident pions on nuclear targets
\cite{Aitala:2001hc} has recently provided a remarkable
confirmation of color
transparency and thus the gauge theory interactions predicted by QCD.
The measured longitudinal momentum distribution of the jets
\cite{Aitala:2001hb} is consistent with a pion light-front wavefunction of
the pion with the shape of the asymptotic distribution amplitude,
$\phi^{\rm asympt}_\pi (x) = \sqrt 3 f_\pi x(1-x)$.  Data from CLEO
\cite{Gronberg:1998fj} for the $\gamma
\gamma^* \rightarrow \pi^0$ transition form factor also favor a
form for the pion distribution amplitude close to the asymptotic
solution to the perturbative QCD evolution equation
\cite{Lepage:1980fj}.
The new EVA spectrometer experiment E850 at
Brookhaven~\cite{Leksanov:2001ui} has also reported striking
effects of color transparency in quasi-elastic large momentum
transfer proton-proton scattering in nuclei which at large
momentum transfer is controlled by proton Fock states of small
transverse size.  The EVA experiment also finds a breakdown of
color transparency at $\sqrt s \sim 5 $ GeV and large
$\theta_{cm}$, which may reflect effects of the charm
threshold~\cite{Brodsky:1987xw}.

\section{THE ANOMALOUS GRAVITOMAGNETIC MOMENT,
ANGULAR MOMENTUM CONSERVATION, AND LIGHT-FRONT WAVEFUNCTIONS}

One can also express the matrix elements of the energy momentum
tensor as overlap integrals of light-front
wavefunctions~\cite{Brodsky:2001ii}.  An important consistency
check of any relativistic formalism is to verify the vanishing of
the anomalous gravito-magnetic moment $B(0)$, the spin-flip matrix
element of the graviton coupling and analog of the anomalous
magnetic moment $F_2(0)$.  For example, at one-loop order in QED,
$B_f(0) = {\alpha \over 3 \pi}$ for the electron when the graviton
interacts with the fermion line, and $B_\gamma(0) = -{\alpha \over
3 \pi}$ when the graviton interacts with the exchanged photon. The
vanishing of $B(0)$ can be shown to be exact for bound or
elementary systems in the light-front
formalism~\cite{Brodsky:2001ii}, in agreement with the equivalence
principle~\cite{Okun}.

The light-front formalism also provides a simple representation of
angular momentum.  See for example,
Refs.~\cite{Ji:1996ek,Brodsky:1997de,%
Harindranath:1999ve,Brodsky:2001ii,Krassnigg:2001ka}.  The
projection $J_z$ is kinematical and is conserved separately for
each Fock component: each light-front Fock wavefunction satisfies
the sum rule: $ J^z = \sum^n_{i=1} S^z_i + \sum^{n-1}_{j=1} l^z_j
\ .  $ The sum over $S^z_i$ represents the contribution of the
intrinsic spins of the $n$ Fock state constituents.  The sum over
orbital angular momenta
$
l^z_j = -{\mathrm i} \left(k^1_j\frac{\partial}{\partial k^2_j}
-k^2_j\frac{\partial}{\partial k^1_j}\right)
$
derives from
the $n-1$ relative momenta.  This excludes the contribution to the
orbital angular momentum due to the motion of the center of mass,
which is not an intrinsic property of the
hadron~\cite{Brodsky:2001ii}.  The numerator structure of the
light-front wavefunctions in $k_\perp$ is determined by the
angular momentum constraints.
The spin properties of light-front wavefunctions
provides a consistent basis for analyzing spin correlations
and azimuthal spin
asymmetries in both exclusive and inclusive reactions.

\section{THE ROLE OF HIGHER PARTICLE-NUMBER FOCK STATES}

The higher Fock states of the light hadrons describe the sea quark
structure of the deep inelastic structure functions, including
``intrinsic" strangeness and charm fluctuations specific to the
hadron's structure rather than gluon substructure
\cite{Brodsky:1980pb,Harris:1996jx}.  Ladder relations connecting
state of different particle number follow from the QCD equation of
motion and lead to Regge behavior of the quark and gluon
distributions at $x \to 0$ \cite{Antonuccio:1997tw}.

Since the intrinsic heavy quarks tend to have the
same rapidity as that of the projectile, they are produced at
large $x_F$ in the beam fragmentation region.  The charm structure
function measured by the EMC group shows an excess at large
$x_{bj}$, indicating a probability of order $1\%$ for intrinsic
charm in the proton~\cite{Harris:1996jx}.  The presence of
intrinsic charm in light-mesons provides an explanation for the
puzzle of the large $J/\psi \to \rho\pi$ branching ratio and
suppressed $\psi^\prime \to \rho\pi$ decay~\cite{Brodsky:1997fj}.
The presence of intrinsic charm quarks in the $B$ wave function
provides new mechanisms for $B$ decay.  For example, Chang and
Hou have considered the production of final states with three
charmed quarks such as $B \to J/\psi D \pi$ and $B \to J/\psi
D^*$~\cite{Chang:2001iy}; these final states are difficult to
realize in the valence model, yet they occur naturally when the
$b$ quark of the intrinsic charm Fock state $\ket{ b \bar u c \bar
c}$ decays via $b \to c \bar u d$.  In fact, the $J/\psi$ spectrum
for inclusive $B \to J/\psi X$ decays measured by CLEO and Belle
shows a distinct enhancement at the low $J/\psi$ momentum where
such decays would kinematically occur.  Alternatively, this excess
could reflect the opening of baryonic channels such as $B \to
J/\psi \bar p \Lambda$~\cite{Brodsky:1997yr}.  Recently, Susan
Gardner and I have shown that the presence of intrinsic charm in
the hadrons' light-front wave functions, even at a few percent
level, provides new, competitive decay mechanisms for $B$ decays
which are nominally CKM-suppressed~\cite{Brodsky:2001yt}.  For
example, the weak decays of the $B$-meson to two-body exclusive
states consisting of strange plus light hadrons, such as $B \to
\pi K$, are expected to be dominated by penguin contributions
since the tree-level $b\to s u{\overline u}$ decay is CKM
suppressed.  However, higher Fock states in the $B$ wave function
containing charm quark pairs can mediate the decay via a
CKM-favored $b\to s c{\overline c}$ tree-level transition.
Since they mimic the amplitude structure of
``charming'' penguin contributions~\cite{Ciuchini:2001gv},
charming penguins need not be penguins at
all~\cite{Brodsky:2001yt}.

\section{GENERAL FEATURES OF LIGHT-FRONT WAVEFUNCTIONS}

Even without explicit solutions, many features of the light-front
wavefunctions follow from general principles and its equation of
motion, Eq.\ (\ref {eq:feqm}). Every light-front Fock wavefunction
has the form:
\begin{equation}
\psi_n={\Gamma_n\over M^2-\sum_{i=1}^n{m_{\perp i}^2\over x_i}}
\end{equation}
where $\Gamma_n = \sum_{n'}\int V_{n {n'}} \psi_n$.  Much
of the dynamical dependence of a light-front wavefunction
is controlled by its light-cone energy denominator.  The
maximum of the wavefunction occurs when the invariant mass of the
partons is minimal; \ie, when all particles have equal rapidity
and are all at rest in the rest frame.  In fact, Dae Sung Hwang
and I \cite{BH2} have noted that one can rewrite the wavefunction
in the form:
\begin{equation} \psi_n= {\Gamma_n\over M^2
[\sum_{i=1}^n {(x_i-{\hat x}_i)^2\over x_i} + \delta^2]}
\end{equation}
where $x_i = {\hat x}_i\equiv{m_{\perp
i}/ \sum_{i=1}^n m_{\perp i}}$ is the condition for minimal
rapidity differences of the constituents.  The key parameter is $
M^2-\sum_{i=1}^n{m_{\perp i}^2/ {\hat x}_i}\equiv -M^2\delta^2.$
We can also interpret $\delta^2 \simeq 2 \epsilon / M $ where $
\epsilon = \sum_{i=1}^n m_{\perp i}-M $ is the effective binding
energy.  This form shows that the wavefunction is a quadratic form
around its maximum, and that the width of the distribution in
$(x_i - \hat x_i)^2$ (where the wavefunction falls to half of its
maximum) is controlled by $x_i \delta^2$ and the transverse
momenta $k_{\perp_i}$.  Note also that the heaviest particles tend
to have the largest $\hat x_i,$ and thus the largest momentum
fraction of the particles in the Fock state, a feature familiar
from the intrinsic charm model.  For example, the $b$ quark has the
largest momentum fraction at small $k_\perp$ in the $B$ meson's
valence light-front wavefunction,, but the distribution spreads out
to an asymptotically symmetric distribution around $x_b \sim 1/2$
when $k_\perp \gg m^2_b.$

We can also discern some general properties of the numerator of
the light-front wavefunctions.  $\Gamma_n(x_i, k_{\perp i},
\lambda_i)$.  The transverse momentum dependence of $\Gamma_n$
guarantees $J_z$ conservation for each Fock state:  For
example, one of the three light-front Fock wavefunctions of a $J_z
= +1/2$ lepton in QED perturbation theory is
\begin{equation}
\psi^{\uparrow}_{+\frac{1}{2}\, +1} (x,{\vec
k}_{\perp})=-{\sqrt{2}} \frac{(-k^1+{\mathrm i} k^2)}{x(1-x)}\,
\varphi \ ,\end{equation}
where
\begin{equation} \varphi (x,{\vec
k}_{\perp})=\frac{ e/\sqrt{1-x}}{\frac{M^2-({\vec
k}_{\perp}^2+m^2)}{x-({\vec k}_{\perp}^2+\lambda^2)/(1-x)}}\ .
\end{equation}
The orbital
angular momentum projection in this case is $\ell^z = -1.$ The
spin structure indicated by perturbative theory provides a
template for the numerator structure of the light-front
wavefunctions even for composite systems.  The structure of the
electron's Fock state in perturbative QED shows that it is natural
to have a negative contribution from relative orbital angular
momentum which balances the $S_z$ of its photon constituents.  We
can also expect a significant orbital contribution to the proton's
$J_z$ since gluons carry roughly half of the proton's momentum,
thus providing insight into the ``spin crisis" in QCD.

The fall-off the light-front wavefunctions at large $k_\perp$ and
$x \to 1$ is dictated by QCD perturbation theory since the state
is far-off the light-cone energy shell.  This leads to counting
rule behavior for the quark and gluon distributions at $x \to 1$.
Notice that $x\to 1$ corresponds to $k^z \to -\infty$ for any
constituent with nonzero mass or transverse momentum.  Explicit examples
of light-front wavefunctions in QED are given
in Ref.~\cite{Brodsky:2001ii}.
The above discussion suggests that an approximate form for the
hadron light-front wavefunctions might be constructed through
variational principles and by minimizing the expectation value of
$H^{QCD}_{LC}.$

\section{THE FORMULATION OF LIGHT-FRONT GAUGE THEORY}

The light-front quantization of QCD in light-cone gauge has a
number of advantages, including explicit unitarity, a physical
Fock expansion, the absence of Faddeev-Popov ghost terms, and the
decoupling properties needed to prove factorization theorems in
high momentum transfer inclusive and exclusive reactions. The
light-front framework for gauge theory in light-cone gauge $n
\cdot A = A^+ = 0,$ is a severely constrained dynamical theory
with many second-class
constraints~\cite{dir1,Nakawaki:1999tc,Srivastava:2000cf}.  These
can be eliminated by constructing Dirac brackets, and the theory
can be quantized canonically by the correspondence principle in
terms of a {\it reduced number} of independent fields.  The
commutation relations among the field operators can be found by
the Dirac method. A recent derivation of light-front quantized
non-Abelian gauge theory in light-cone gauge is given in
Ref.~\cite{Srivastava:2000cf}. For example, the nondynamical
projections of the fermion and gauge field can be eliminated using
nonlocal constraint equations.  The removal of the unphysical
components of the fields results in tree-level instantaneous gluon
exchange and fermion exchange interaction terms.  The interaction
Hamiltonian of QCD can be expressed in a form resembling that of
covariant theory with three- and four-point gauge interactions,
except for the additional instantaneous four-point interactions
which can be treated systematically.

The light-front quantized free gauge theory simultaneously satisfies the
covariant gauge condition $\partial\cdot A=0$ as an operator condition as
well as the light-cone gauge condition.  In our
analysis~\cite{Srivastava:2000cf} one imposes this condition linearly
using a Lagrange multiplier, rather than a quadratic form.
The numerator of the gauge propagator is doubly transverse:
$$
D_{\mu\nu}(k)=-g_{\mu\nu} + \frac
{n_{\mu}k_{\nu}+n_{\nu}k_{\mu}}{(n\cdot k)} - \frac {k^{2}}
{(n\cdot k)^{2}} \, n_{\mu}n_{\nu},$$
with
$$ n^\mu  D_{\mu \nu} = k^\mu  D_{\mu \nu} =0.$$
Thus only
physical degrees of freedom propagate.
The remarkable properties of $D_{\nu\mu}$ provide much
simplification in the computations of loops.
In the case of tree graphs, the
term proportional to
$n_{\mu}n_{\nu}$ cancels against the instantaneous gluon exchange term
leading to Eq.
(\ref{lcprop}).  The renormalization constants in the non-Abelian
theory can be shown to can be shown to satisfy the identity
$Z_1=Z_3$ at one loop order, as expected in a theory with only
physical gauge degrees of freedom.  Note that the one-loop
correction to the three gluon vertex includes an instantaneous
gluon exchange contribution since it is not one-particle
irreducible.  Thorn~\cite{Thorn:1979gv} gave the first
computation of the renormalization constants of QCD in light-cone
gauge. The QCD $\beta$ function computed in the
noncovariant light-cone gauge~\cite{Srivastava:2000cf} agrees with
the conventional result.  Dimensional regularization and the
Mandelstam-Leibbrandt prescription~\cite{Leibbrandt:1987qv} for
light-cone gauge were used to define the Feynman loop integration
\cite{Bassetto:1996ph}. Ghosts only appear in association with the
Mandelstam-Liebbrandt prescription.  There are no Faddeev-Popov or
Gupta-Bleuler ghost terms.

Bassetto, Griguolo, and Vian~\cite{Bassetto:2000qz} have
investigated the presence of multiple vacua and the use of new
prescriptions for light-cone gauge.  In previous work they showed
that in two dimensions (where no UV singularities occur), both the
Mandelstam-Liebbrandt and principal value prescriptions for
light-cone gauge are viable.  In general they lead to different
theories and to different results.  Since pure Yang Mills theory
in 1+1 dimensions is exactly solvable by geometric techniques
without even fixing a gauge, Bassetto et
al.~\cite{Bassetto:1999er,Bassetto:2000qz} could show that the
result of the principal value prescription coincides with the
exact solution, whereas the Mandelstam-Leibbrandt procedure misses
the topological effects.

In an important new development, Paston, Prokhvatilov and Franke,
have shown that one can introduce a Pauli Villars spectrum of
ghost fields to regulate the ultra-violet divergences of
non-Abelian gauge theory~\cite{Paston:2001ik}.  This provides an
important computational tool for the defining nonperturbative QCD
without dimensional regularization.  They have also demonstrated
the equivalence between Light-Front Hamiltonian and conventional
Lorentz-covariant formulations of gauge theory for QED(1+1) and
perturbatively to all orders for QCD(3+1).

One of the most interesting formal questions in light-front
quantization is how solutions of quantum field theories which
display spontaneous symmetry emerges despite the simplicity of the
light-front vacuum~\cite{Pinsky:1994si}.  Spontaneous symmetry
breaking and other nonperturbative effects associated with the
instant-time vacuum are hidden in dynamical or constrained zero
modes on the light-front.  An introduction is given in
Ref.~\cite{McCartor:hj}.  Significant progress in this area was
reported at this meeting.  Grange, Salmons, and
Werner~\cite{Grange:2001kt} have given a new analysis of
spontaneous symmetry breaking of $\phi^4$ theory in 1+1
dimensions.  Pirner~\cite{Pirner:2001pv} has shown how zero modes
on the light-cone in QCD can cause a phase transition which can be
studied phenomenologically in deep inelastic lepton scattering.
The relation of zero modes to the $\theta$ vacuum of the bosonized
version of QCD(1+1) was also discussed at this meeting in talks by
Przeszowski~\cite{Przeszowski:1999gc}
McCartor~\cite{Nakawaki:2000yu}, Martinovic~\cite{Martinovic:va}.

\section{NONPERTURBATIVE CALCULATIONS OF LIGHT-FRONT WAVEFUNCTIONS}

The calculation of the light-front wavefunctions of hadrons from first
principles is a central goal in QCD.  A number of methods have been
proposed, all of which have shown considerable progress.

DLCQ (discretized light-cone quantization) is a method which
solves quantum field theory by directly diagonalization of the
light-front Hamiltonian~\cite{Pauli:1985pv}.  The DLCQ method has
been used with success in solving a number of quantum field
theories in low space-time dimensions~\cite{Gross:1997mx} and has
found much utility in string theory.  Susskind and others have
adopted DLCQ to define $M-$theory
\cite{Susskind:1997cw,Antonuccio:1996cf}.  Harinandrath \etal
~\cite{Harindranath:2001eb} have shown that the $S-$matrix in DLCQ
does have the correct continuum limit.
When one imposes periodic boundary conditions in $x^- = t + z/c$,
then the plus momenta become discrete: $k^+_i = {2\pi \over L}
n_i, P^+ = {2\pi\over L} K$, where $\sum_i n_i =
K$~\cite{Maskawa:1975ky,Pauli:1985pv}.  For a given ``harmonic
resolution" $K$, there are only a finite number of ways positive
integers $n_i$ can sum to a positive integer $K$.  Thus at a given
$K$, the dimension of the resulting light-front Fock state
representation of the bound state is rendered finite without
violating Lorentz invariance.  The eigensolutions of a quantum
field theory, both the bound states and continuum solutions, can
then be found by numerically diagonalizing a frame-independent
light-front Hamiltonian $H_{LC}$ on a finite and discrete
momentum-space Fock basis.  Solving a quantum field theory at
fixed light-front time $\tau$ thus can be formulated as a
relativistic extension of Heisenberg's matrix mechanics.  The
continuum limit is reached for $K \to \infty.$ This formulation of
the non-perturbative light-front quantization problem is called
``discretized light-front quantization" (DLCQ)~\cite{Pauli:1985pv}.
Lattice gauge theory has also been used to calculate the pion
light-front wavefunction~\cite{Abada:2001if}.

\begin{figure*}
{\includegraphics[height=3in,width=6in]{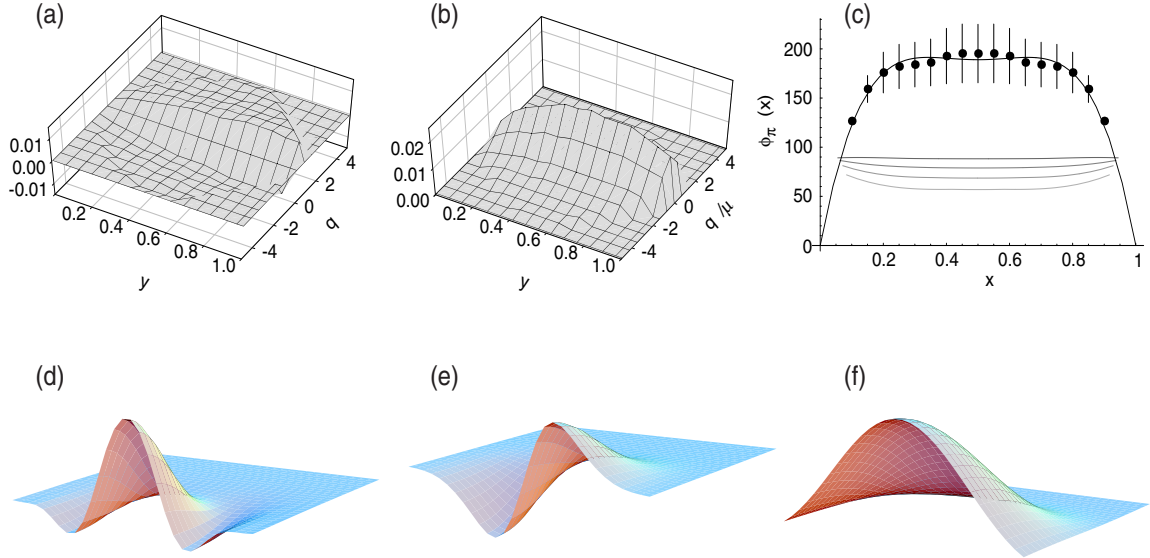}} 
\caption[*]{\label{fig:phi1000} DLCQ results for the one-boson
one-fermion wavefunction in a fermion system with (a) parallel and
(b) antiparallel fermion helicity, as a function of longitudinal
momentum fraction $y$ and one transverse momentum component $q_x$
in the $q_y=0$ plane.  The parameter values for the DLCQ
resolution are $K=29$, $N_\perp=7$.  Further details are given in
Ref.~\cite{Brodsky:2001ja}. (c) The distribution amplitude
$\phi(x)$ of the pion for opposite helicity quarks computed from
the transverse lattice by Dalley~\cite{Dalley:2001gj} at a
transverse normalization scale $Q^2 \sim 1$ $ {\rm GeV}^2$.  Grey
curves correspond to DLCQ cut-offs $K=6,7,8,9$ (darker means
larger $K$).  Data points are the point-wise extrapolation of
finite-$K$ curves.  The solid curve is a fit of the distribution
amplitude to the conformal expansion. (d)-(e) Light-front
wavefunctions for the first three eigensolutions of an effective
3+1 light-front equation generated from $\phi^3$ field
theory~\cite{vanIsrael}.  The vertical axis shows the value of the
two-parton wavefunction.  The front of the figure shows
$\psi(x,\vec k_\perp = 0)$.  (d) The ground state wavefunction
peaks at $x = 1/2$ falling to zero $x = 0$ and $x =1.$ (e) The
first excited state has a node in $\psi(x, \vec k_\perp = 0)$ at
$x = 1/2.$ The figure also shows the fall-off of the wavefunctions
as a function of the magnitude of the transverse momentum as one
goes from front to back.  Further details are given in
Ref.~\cite{vanIsrael}.
 }
\end{figure*}

The DLCQ method has been used extensively for solving one-space
and one-time theories~\cite{Brodsky:1997de}, including
applications to supersymmetric quantum field
theories~\cite{Matsumura:1995kw} and specific tests of the
Maldacena conjecture~\cite{Hiller:2001mh}.  There has been
progress in systematically developing the computation and
renormalization methods needed to make DLCQ viable for QCD in
physical spacetime. For example, John Hiller, Gary McCartor, and
I~\cite{Brodsky:2001ja} have shown how DLCQ can be used to solve
3+1 theories despite the large numbers of degrees of freedom
needed to enumerate the Fock basis.  A key feature of our work is
the introduction of Pauli Villars fields to regulate the
ultraviolet divergences and perform renormalization while
preserving the frame-independence of the theory.  A recent
application of DLCQ to a 3+1 quantum field theory with Yukawa
interactions is given in Ref.~\cite{Brodsky:2001ja}.
Representative plots of the one-boson one-fermion light-front Fock
wavefunction of the lowest mass fermion solution of the Yukawa
(3+1) theory showing spin correlations and the presence of
non-zero orbital angular momentum are shown in
Fig.~\ref{fig:phi1000}(a),(b). Hiller, McCartor and I have also
shown that one can obtain exact analytic DLCQ solutions when the
Pauli-Villars and normal particles are degenerate. This is
valuable for checking DLCQ solutions and
codes~\cite{Brodsky:2001tp}.

Pinsky, Trittman and Hiller~\cite{Hiller:2001mh}
have shown how one can solve supersymmetric
gauge theories by discretizing and diagonalizing the $Q$ operators of
DLCQ.  The square of $Q$ then provides $H_{LC}$.  They have also used
SDLCQ to give explicit checks of the Maldacena conjecture.

Dyson-Schwinger solutions~\cite{Hecht:2000xa} of hadronic
Bethe-Salpeter wavefunctions can also be used to predict
light-front wavefunctions and hadron distribution amplitudes by
integrating over the relative $k^-$ momentum.
Roberts~{\etal}~\cite{Hecht:2000xa} have made important progress
treating meson states in QCD in terms of a Bethe-Salpeter/
Dyson-Schwinger approach.  Much of the physics of chiral symmetry
breaking can be simulated in this formalism without an explicit
vacuum chiral condensate.  A disadvantage of this approach is that
the ladder approximation to the Bethe-Salpeter kernel is gauge
dependent and breaks crossing symmetry.  In addition, covariant
gauges with ghost terms are required.

There has been important progress using the transverse
lattice, essentially a combination of DLCQ in 1+1 dimensions
together with a lattice in the transverse
dimensions~\cite{Bardeen:1979xx,Dalley:2001gj,Burkardt:2001dy}.  For
example, Dalley and Burkardt have used the transverse lattice to compute
the Isgur-Weis function measured in $B \to D \ell \bar \nu$ decays.
Dalley \cite{Dalley:2000dh} has also calculated the pion
distribution amplitude from QCD using this method.  A finite lattice
spacing $a$ can be used by choosing
the parameters of the effective theory in a region of
renormalization group stability to respect the required gauge,
Poincar\'e, chiral, and continuum symmetries.  The overall normalization gives
$f_{\pi} = 101$ MeV compared with the experimental value of $93$ MeV.
An illustration is shown in Fig. \ref{fig:phi1000}(c).
The
resulting DLCQ/transverse lattice pion wavefunction compares well
with the best fit to the
diffractive di-jet data after corrections for hadronization and experimental
acceptance \cite{Ashery:1999nq}.  The predicted form of
$\phi_\pi(x,Q)$ is somewhat broader than but not inconsistent with the
asymptotic form favored by the measured normalization of $Q^2 F_{\gamma
\pi^0}(Q^2)$ and the pion wavefunction inferred from diffractive di-jet
production.  However, there are experimental uncertainties from hadronization
and theoretical errors introduced from finite DLCQ resolution, using a nearly
massless pion, ambiguities in setting the factorization scale
$Q^2$, as well as errors in the evolution of the distribution amplitude from 1
to $10~{\rm GeV}^2$.  Instanton models also predict a pion distribution
amplitude close to the asymptotic form~\cite{Petrov:1999kg}.

Moments of the pion distribution amplitudes have been also
computed in Euclidean lattice gauge theory
\cite{Martinelli:1987si,DelDebbio:2000mq}.  The lattice results
from Del Debbio {\em et al.}~\cite{DelDebbio:2000mq} imply a much
narrower shape for the pion distribution amplitude than the
asymptotic form or the distribution predicted by the transverse
lattice.  A new result for the proton distribution amplitude
treating nucleons as chiral solitons has recently been derived by
Diakonov and Petrov~\cite{Diakonov:2000pa}.

A systematic approach to QCD has been developed by Wilson
\etal~\cite{Wilson:1994fk} and others which systematically
eliminates higher Fock states in terms of new renormalization
constants and effective interactions.  Recent progress on the
dynamics of effective gluons, renormalization, and the $k^+ \to 0$
behavior of the effective theory has been reported by Glazek
\cite{Glazek:2001gb}, Brisudova~\cite{Brisudova:2001pb} and
Walhout~\cite{Barnea:nu}.
One can also define a truncated theory by eliminating the higher
Fock states in favor of an effective
potential.~\cite{Coester:2001eb} For example Pauli, {\it et
al.}~\cite{Pauli:2001vi} and Karmanov \etal,~\cite{Bernard:2001id}
have shown how one can develop effective potentials which in
principle can systematically incorporates the effects of higher
Fock states. New methods have been developed by Mangin-Brinet
\etal~\cite{Mangin-Brinet:ma} and Fredirico
\etal,~\cite{Frederico:2001rr} to solve the effective equations.

van Iersel, Bakker, and Pijlman \cite{vanIsrael} have shown how
one can obtain bound state solutions and explicit light-front
wavefunctions for theories with simple (3+1) field theory
interactions.  As an example, they consider a system consisting of
two scalar particles of equal mass $m =1$ exchanging a massless
spinless particle, allowing for 2- and 3-particle Fock states.  The
$g\phi^3$ coupling strength is $\alpha = {g^2 \over 16 \pi m^2} =
17.36.$ They solve the light-front bound state equation by making
an expansion in basis functions -- 8 cubic spline functions for
the functional dependence in $x$ and up to 3 Jacobi polynomials in
the transverse momenta.  Figure~\ref{fig:phi1000}(d)-(e) shows the valence
wavefunctions of the first three bound states.  It is interesting
to see how the node structure appears in the light-cone fraction
$x$ in the higher excited states.

There are other possible approaches to non-perturbative QCD using
light-front quantization.  For example, one can construct trial
wavefunctions for the light-front wavefunctions which incorporate
ladder relations between Fock states, angular momentum
constraints, and perturbative QCD fall-off at high off-shell
virtuality.  The parameters would be determined variationally by
minimizing the expectation value of $H_{LC}.$
Lattice theory could also be adopted to light-cone coordinates
$x^+,x^-,x_\perp$ rather than Cartesian
coordinates, thus incorporating the good Lorentz boost properties of
light-front quantization.

\section{NEW DIRECTIONS}

Light-cone quantization in light-cone gauge can be applied to the
electroweak theory, including spontaneous symmetry breaking, thus
providing a unitary as well as renormalizable theory of the
Standard Model~\cite{SB}.

The light-front formalism can be used as an ``event amplitude
generator" for high energy physics reactions where each particle's
final state is completely labeled in momentum, helicity, and
phase~\cite{Brodsky:2001ww}. The application of the light-front
time evolution operator $P^-$ to an initial state systematically
generates the tree and virtual loop graphs of the $T$-matrix in
light-front time-ordered perturbation theory in light-front gauge.
Renormalized amplitudes can be explicitly constructed by
subtracting from the divergent loops amplitudes with nearly
identical integrands corresponding to the contribution of the
relevant mass and coupling counter terms (the ``alternating
denominator method")~\cite{Brodsky:1973kb}.

In the usual treatment of classical thermodynamics, one considers
an ensemble of particles $n = 1, 2, \ldots N$ which have energies
$\{E_n\}$ at a given ``instant" time $t$.  The partition function
is defined as $Z = \sum_n \exp-{E_n\over kT}.$ Similarly, in
quantum mechanics, one defines a quantum-statistical partition
function as $Z = tr \exp{-\beta H}$ which sums over the
exponentiated-weighted energy eigenvalues of the system.
In the case of relativistic systems, it is natural to characterize
the system at a given light-front time $\tau = t +z/c$; {\em
i.e.}, one determines the state of each particle in the ensemble
as its encounters the light-front.  Thus we can define a
light-front partition function~\cite{Brodsky:2001ww}
$Z_{LC} = \sum_n \exp -{p^-_n\over kT_{LC}}$
by
summing over the particles' light-front energies $p^- = p^0 - p^z
= {p^2_\perp + m^2 \over p^+}$.  The total momentum is $P^+ = \sum
p^+_n,$ $ \vec P_\perp = \sum_n \vec p_{\perp n}$, and the total
mass is defined from $P^+P^--P^2_\perp=M^2$.  The light-front
partition function should be advantageous for analyzing
relativistic systems such as heavy ion collisions, since, like
true rapidity, $y = \ell n {p^+\over P^+},$ light-front variables
have simple behavior under Lorentz boosts.  The light-front
formalism also takes into account the point that a phase
transition does not occur simultaneously in $t$, but propagates
through the system with a finite wave velocity.

\section{Conclusions}

There has been strong advances in light-cone physics, particularly
in solving many formal problems of renormalization and implementation of
spontaneous symmetry breaking within light-front quantized field theory.
There have also been important new developments using light-cone methods
in string theory and supersymmetry.  The Pauli-Villars method
has now emerged as a systematic way to regulate ultraviolet divergences
in light-front quantized non-Abelian gauge theory.
The light-front formalism thus can provide a formalism for solving
nonperturbative problems in QCD as rigorous as lattice gauge
theory.

On the phenomenological side, the measurements from the Fermilab E791 experiment
of diffractive dijet production in pion-nucleus collisions has not only
verified color transparency, but also given us a direct look
at the light-front wavefunction of the pion.  It will be important to extend
these measurements to the diffractive dissociation of real and virtual
photons, nucleons, and even light nuclei.

The new applications of light-front wavefunctions to exclusive $B$ decays
makes even more urgent the need for bound state solutions of the light-front
Hamiltonian for QCD in physical space-time dimensions.  Strong progress has
been made
solving model $3+1$ field theories using DLCQ,
light-front Tamm-Dancoff,
effective light-front potential equations, and the Dyson-Schwinger approach.
The transverse lattice method is now providing the first results
for the pion distribution amplitude and the
Isgur-Weis function for heavy hadron decays from QCD.  Variational methods
also look promising.  Given these initial successes
in (3+1) studies,  we can be optimistic that with sufficient computational
resources, first principle calculations of the spectrum and light-front
wavefunctions of hadrons
from light-front quantized QCD will soon emerge.

\section{Acknowledgment}
I thank Antonio Bassetto, Federica Vian, and the ECT* for organizing and
hosting this outstanding ILCAC meeting.  I also thank Steven Bass,
Markus Diehl,  Susan Gardner, Dae Sung
Hwang, John Hiller, Paul Hoyer, Bo-Qiang Ma, Nils Marchal, Gary McCartor,
Stephane Peigne, Hans Christian Pauli, Steve Pinsky, Francesco Sannino, Ivan
Schmidt, Prem Srivastava, Charles Thorn, and Miranda van Iersel for their
valuable
input.  This work was supported by the Department of Energy under contract
number DE-AC03-76SF00515.

\end{document}